\documentclass[reprint,groupedaddress,amsmath,amssymb,aps]{revtex4-2}

\usepackage{graphicx}
\usepackage{dcolumn}
\usepackage{bm}
\usepackage{empheq}
\usepackage{amsmath}
\usepackage{enumitem}
\usepackage{xcolor}

\begin{document}

\preprint{APS/123-QED}

\title{Playing with Active Matter}

\author{Angelo Barona Balda}
\affiliation{Department of Physics, University of Gothenburg, Gothenburg, Sweden}

\author{Aykut Argun}
\affiliation{Department of Physics, University of Gothenburg, Gothenburg, Sweden}

\author{Agnese Callegari}
\affiliation{Department of Physics, University of Gothenburg, Gothenburg, Sweden}

\author{Giovanni Volpe}
\email{giovanni.volpe@physics.gu.se}
\affiliation{Department of Physics, University of Gothenburg, Gothenburg, Sweden}

\date{\today}

\begin{abstract}
In the last 20 years, active matter has been a very successful research field, bridging the fundamental physics of nonequilibrium thermodynamics with applications in robotics, biology, and medicine. 
This field deals with active particles, which, differently from passive Brownian particles, can harness energy to generate complex motions and emerging behaviors.
Most active-matter experiments are performed with microscopic particles and require advanced microfabrication and microscopy techniques.
Here, we propose some macroscopic experiments with active matter employing commercially available toy robots, i.e., the Hexbugs.
We demonstrate how they can be easily modified to perform regular and chiral active Brownian motion.
We also show that Hexbugs can interact with passive objects present in their environment and, depending on their shape, set them in motion and rotation.
Furthermore, we show that, by introducing obstacles in the environment, we can sort the robots based on their motility and chirality.
Finally, we demonstrate the emergence of Casimir-like activity-induced attraction between planar objects in the presence of active particles in the environment.
\end{abstract}


\maketitle

\begin{figure*}
    \includegraphics[width=\linewidth]{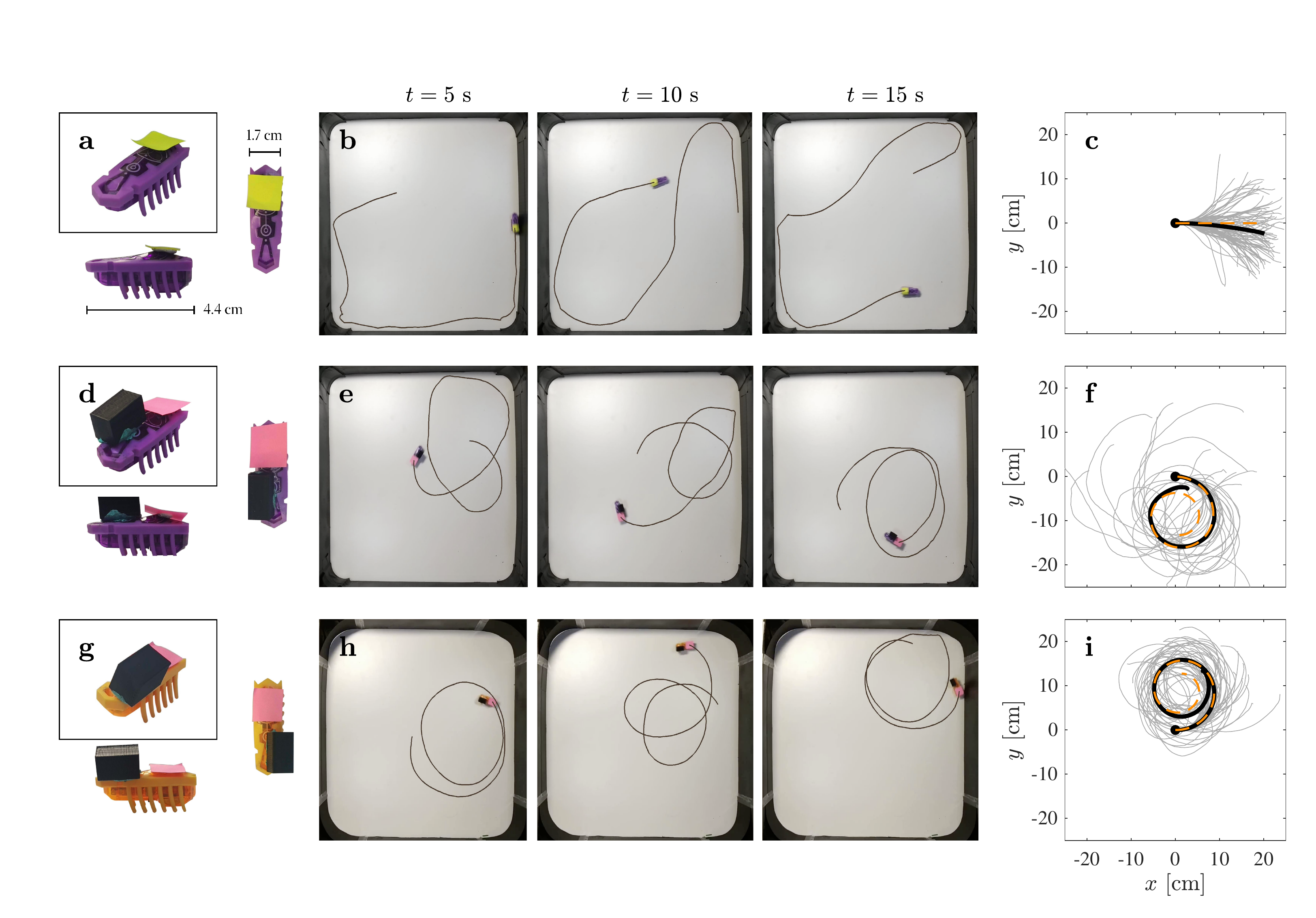}
    \caption{
    {\bf Hexbugs perform regular and chiral active Brownian motion.}
    {\bf a} Pictures of an Hexbug with a yellow tag to identify it.
    {\bf b} Its motion is a regular active Brownian motion, as can be seen by its trajectories in a square arena ($44\,{\rm cm} \times 44 \,{\rm cm}$) with rounded corners (see Video~1).
    {\bf c} This motion can be quantified by calculating the average trajectory in the frame of reference of the Hexbug (thick black line; the gray lines show the 91 trajectories used in the averaging, selected so that the robot does not interact with the boundaries of the arena).
    {\bf d} By adding a weight (black 3D-printed plastic parallelepiped) to the Hexbug, it is possible to control its motion. 
    {\bf e} When the weight is placed on the Hexbug starbord side, the robot performs a right-chiral (dextrogyre) active Brownian motion (see Video~2).
    {\bf f} The average trajectory in the Hexbug frame of reference (thick black line) clearly bends rightward, which can be fitted to a {\it spira mirabilis} (orange line). (The 22 gray trajectories are used in the averaging.)
    {\bf g} By placing the weight on the Hexbug port side, {\bf h} it performs a left-chiral (levogyre) active Brownian motion (see Video~3), {\bf j} whose average trajectory (thick solid line, obtained averaging the 29 gray trajectories) and {\it spira mirabililis} (orange line) bend leftward.
    }
    \label{fig:1}
\end{figure*}

\section{Introduction}

Active-matter systems consist of ``active particles'', which are living or synthetic agents that generate mechanical forces resulting in self-propulsion \cite{romanczuk2012active,bechinger2016active}. 
Some typical examples of active motion include the swimming behavior of {\it E. coli} bacteria \cite{berg2008coli} and the self-propulsion of artificial Janus particles \cite{paxton2005motility,valadares2010catalytic,buttinoni2012active}.

Active-matter systems occur at all length scales \cite{yeomans2017nature}, ranging from microscopic colloidal systems \cite{palacci2013living,bechinger2016active} to groups of animals or robots \cite{cavagna2010scalefree,rubenstein2014kilobot} and even human crowds \cite{helbing1995social,warren2018collective}. 
In spite of their seemingly large differences, all these systems show similar emergent behaviors. 
Important examples are the motion observed in flocking birds \cite{bialek2012statistical}, the formation of living crystals made of active particles  \cite{palacci2013living}, the motility-induced phase separation seen in suspensions of self-propelled hard spheres \cite{stenhammar2013continuum}, the chaotic swarming observed in swimming bacteria \cite{stenhammar2017role}, and the formation of fruiting bodies in {\it M. xanthus} bacterial colonies \cite{ramos2021environment}.

Performing experiments with active particles can provide key insights into nonequilibrium physics. 
In fact, active motion leads to a steady dissipation of energy into the environment. This makes active particles intrinsically out of thermodynamic equilibrium \cite{ramaswamy2010mechanics}.
By being out of equilibrium, active particles can dynamically self-organize and exhibit collective behaviors that are impossible at thermodynamic equilibrium \cite{bechinger2016active}.

Performing experiments can also reveal how active matter might benefit technological applications. Some examples are delivering drugs to target organs \cite{ghosh2020active}, controlling the spread of infectious microorganisms \cite{forgacs2022using}, and developing microrobots capable of advanced group behaviors \cite{rubenstein2014kilobot,yigit2019programmable}.
Most active-matter experiments are, however,
performed with microscopic particles and require advanced microfabrication and microscopy techniques \cite{love2002fabrication}.

In this article, we introduce a simple macroscopic experimental model for active matter based on some commercially available toy robots, i.e., the Hexbugs \cite{hexbug}. Hexbugs are small robots that self-propel by vibrating without control over their direction, resulting in a random pattern of motion.
After showing that this motion can be described as an \emph{active Brownian motion}, we demonstrate how it can be turned into a \emph{chiral} active Brownian motion by some small modificaitons to the robots.
Then, we make the Hexbugs interact with different obstacles. We show how they can propel and rotate some simple passive objects. We also make them interact with complex environments, where they are sorted based on their motility and chirality. 
Finally, we demonstrate the emergence of Casimir-like activity-induced attraction between planar objects in the presence of active particles in the environment.

\section{A Hexbug as an active particle}

Let us consider a single active particle moving in a two-dimensional space. 
Its position is given by the coordinates $[x(t), y(t)]$.
Its self-propulsion results in a directed motion with speed $v$ along a direction that depends on the particle orientation $\varphi(t)$ (defined with respect to the $x$-axis), which undergoes rotational diffusion with a rotational diffusion coefficient $D_{\rm R}$.
For example, the described particle could be a motile bacterium \cite{berg2008coli} or a Janus particle \cite{buttinoni2012active} propelling itself while under the influence of rotational Brownian motion. It could also be a small vibrational robot, like a Hexbug (see Video~1). 

A Hexbug is shown in Fig.~\ref{fig:1}a.
It has six curved rubber legs on each side and a vibrational motor powered by an LR44 battery.
When turned on, the vibration of the motor propagates to the legs. The resulting friction between the legs and the surface causes the robot to move forward. Due to noise and small imperfections in both the terrain and the robot, the orientation of the Hexbug changes over time, leading to rotational diffusion.

To record its motion, we allowed a Hexbug to move within a $44\,{\rm cm}\times44\,{\rm cm}$ arena bounded by some carton walls, as shown in Fig.~\ref{fig:1}b.
We placed a camera above the table to film the Hexbug in motion. We glued a bright-colored tag to the robot and tracked it using the OpenCV library (this library is available for Python and C++ and has several features for computer vision, image processing, and video analysis \cite{opencv}).

In the next subsections, we characterize the motion of a single Hexbug and show that it is in fact an active Brownian motion. Then, we show how a Hexbug can be modified so that it performs a \emph{chiral} active Brownian motion.

\subsection{Active Brownian motion}

The equations describing the active Brownian motion of a single particle are \cite{volpe2014simulation}:
\begin{eqnarray}
    \frac{d}{dt}\varphi(t) 
    &=& 
    \sqrt{2D_{\rm R}}W_{\phi}, \label{eq:activephi}\\
    \frac{d}{dt}x(t)
    &=& 
    v\cos\varphi(t), \label{eq:activex}\\
    \frac{d}{dt}y(t)
    &=& v\sin\varphi(t), \label{eq:activey}
\end{eqnarray}
where $W_\phi$ is a stochastic process with a mean of 0 and a variance of 1.
(Note that for microscopic particles subject to Brownian motion, also translational diffusion terms are present \cite{volpe2014simulation}, but we do not need them in this work with macroscopic robots so we have not included them in the equations.)

We estimate $v$ from the experimental trajectories of the Hexbug by calculating the average speed as
\begin{equation}
    v = \left\langle \frac{ \Delta r }{\Delta t} \right\rangle,
    \label{eq:propulsionspeed}
\end{equation}
where $\Delta r$ is the magnitude ($\ge0$) of the distance covered by the Hexbug during each time interval $\Delta t$. For the trajectory shown in Fig.~\ref{fig:1}b, we estimate $v = 25.7\,{\rm cm\,s^{-1}}$.
Similarly, we estimate $D_{\rm R}$ by calculating the mean of the square of the change in orientation as
\begin{equation}
    D_{\rm R} = \left\langle \frac{\left( \Delta \phi \right)^2}{2 \Delta t} \right\rangle,
    \label{eq:rotationaldiffusion}
\end{equation}
where $\Delta \phi$ is the Hexbug orientation change during $\Delta t$. (Note that  $\langle \Delta \phi \rangle = 0$, because the particle has no preferred reorientation direction.)
For the trajectory shown in Fig.~\ref{fig:1}b, we estimate $D_{\rm R}= 0.21\,{\rm rad^2\, s^{-1}}$.

The ensuing motion is characterized by a persistence length $L$. This is a characteristic travelled distance before rotational diffusion alters the direction of motion of the particle. It is given by
\begin{equation}
    L = \frac{v}{D_{\rm R}}.
    \label{eq:persistencelength}
\end{equation}
The persistence length can be determined by measuring the average motion of the Hexbug in its frame of reference \cite{bechinger2016active,zottl2016emergent}.
We do that by: 
(1) splitting the trajectory into smaller parts that represent the motion between collisions with the walls; 
(2) translating and rotating each part of the trajectory so that all of them start at $x=0$, $y=0$, and $\phi=0$ (gray lines in Fig.~\ref{fig:1}c); 
(3) averaging all these splits.
The resulting average trajectory is
\begin{eqnarray}
    \langle x(t) \rangle 
    &=& 
    \frac{v}{D_{\rm R}}\left[1-e^{-D_{\rm R}t} \right],
    \label{eq:averagex}\\
    \langle y(t) \rangle 
    & \equiv &
    0,
    \label{eq:averagey}
\end{eqnarray}
where $\langle x(t) \rangle$ converges to $L$ for $t \to \infty$.
The resulting mean trajectory is shown by the thick black line in Fig.~\ref{fig:1}c.
Combining the estimations of $v$ and $D_{\rm R}$, we obtain the estimation of the persistence length $L = 122\, {\rm cm}$, which is much longer that the arena where the Hexbug moves.

\subsection{Chiral active Brownian motion}

We can transform a Hexbug into a \emph{chiral} active Brownian particle by gluing a small 3D-printed parallelepiped on top of the robot. 
If this object is glued in an asymmetric position, its weight continuously exerts a torque that bends the trajectory in a consistent direction.
For example, when the weight is placed on the Hexbug starboard (right-hand) side, as shown in Fig.~\ref{fig:1}d, the robot tends to turn right. This leads to a right-chiral  active Brownian motion.
The resulting motion is shown in Fig.~\ref{fig:1}e (Video~2).
By placing the weight on the Hexbug port (left-hand) side (Fig.~\ref{fig:1}g), the robot tends to turn left leading to a left-chiral  active Brownian motion (Fig.~\ref{fig:1}h and Video~3).
The impact of the added weight on the motion chirality might slightly vary for each Hexbug due to some small imperfections, but this can be compensated by finely tuning the exact position of the weight.

The resulting motion can be described by adding an angular velocity $\Omega$ to the equation for the orientation \cite{volpe2014simulation}, obtaining
\begin{eqnarray}
    \frac{d}{dt}\varphi(t) 
    &=& 
    \Omega + 
    \sqrt{2D_{\rm R}}W_{\phi}, \label{eq:chiralphi}
    \\
    \frac{d}{dt}x(t)
    &=& 
    v\cos\varphi(t),  \label{eq:chiralx}\\
    \frac{d}{dt}y(t)
    &=& v\sin\varphi(t)  \label{eq:chiraly}.
\end{eqnarray}
The sign of $\Omega$ defines the orientation of the rotation of the robot. 

We estimate the speed $v$ using Eq.~\eqref{eq:propulsionspeed}, $\Omega$ by calculating the average orientation change as
\begin{equation}
    \Omega = \left\langle \frac{\Delta \phi }{\Delta t} \right\rangle,
    \label{eq:angularvelocity}
\end{equation}
and $D_{\rm R}$ by calculating 
\begin{equation}
    D_{\rm R} = \left\langle \frac{\left( \Delta \phi - \Omega \Delta t \right)^2}{2 \Delta t} \right\rangle,
    \label{eq:chiralrotationaldiffusion}
\end{equation}
which generalizes Eq.~\ref{eq:rotationaldiffusion} to take into account the presence of an angular drift.
We obtain $v= 21.2\,{\rm cm\,s^{-1}}$, $\Omega = -2.16\,{\rm rad\,s^{-1}}$, and $D_{\rm R} = 0.115\, {\rm rad^2\,s^{-1}}$ for the right-chiral Hexbug (Figs.~\ref{fig:1}d-f and Video~2), and $v= 22.0\, {\rm cm\,s^{-1}}$, $\Omega= 2.70\, {\rm rad\,s^{-1}}$, and $D_{\rm R}= 0.061\, {\rm rad^2\,s^{-1}}$ for the left-chiral Hexbug (Figs.~\ref{fig:1}g-i and Video~3).

Also in the chiral case, we can measure the average motion of the Hexbug in its frame of reference.
The resulting average trajectory (thick black lines in Figs.~\ref{fig:1}f and \ref{fig:1}i) can be described by a {\it spira mirabilis} \cite{teeffelen2008dynamics}:
\begin{eqnarray}
    \langle \Delta x(t) \rangle 
    &=& \displaystyle
    L \left\lbrack \cos{\left( \alpha - \theta_0 \right)} - e^{-D_{\rm R} t} \cos{\left( \alpha - \overline{\theta(t)} \right)} \right\rbrack, \label{eq:chiraldx}
    \\[10pt]
    \langle \Delta y(t) \rangle 
    &=& \displaystyle
    L \left\lbrack \sin{\left( \alpha - \theta_0 \right)} - e^{-D_{\rm R} t} \sin{\left( \alpha - \overline{\theta(t)} \right)} \right\rbrack, \label{eq:chiraldy}
\end{eqnarray}
where 
\begin{equation}
    L=\frac{v}{\sqrt{ D_{\rm R}^2+\Omega^2 }}
    \label{eq:persistencechiral}
\end{equation}
is the chiral active particle persistence length, $\alpha$ is the pitch angle of the {\em spira mirabilis} ($\tan{\alpha} = D_{\rm R}/\left|\Omega\right|$), $\theta_0$ is the angle of the initial point of the trajectories with respect to the center $C$ of the {\em spira mirabilis}, and $\overline{\theta(t)} = \theta_0 + \Omega t$ is the angle of the point $(\langle \Delta x(t) \rangle , \langle \Delta y(t) \rangle )$ with respect to $C$. 
The independent parameters determining the \textit{spira mirabilis} are $v$, $\Omega$, and $D_{\rm R}$. Using the estimated values, we obtain the \textit{spira mirabilis} for the right-chiral and left-chiral Hexbug, shown by the orange dashed lines in Figs.~\ref{fig:1}f and ~\ref{fig:1}i, respectively, which fit well with the experimentally measured ones (solid black lines).
We also obtain the persistence length $L = 9.81\, {\rm cm}$ for the right-chiral Hexbug and $L = 8.15\, {\rm cm}$ for the left-chiral Hexbug.
(It is worth noting that, when expressed in the coordinates of its center $C$, a {\em spira mirabilis} is described by the simple dependence $\rho = \rho_{0} e^{a \theta}$ where $\rho$ and $\theta$ are the polar coordinates of one of its points, and $\rho_{0}$ and $a$ are its defining parameters.)

\begin{figure*}
    \includegraphics[width=\linewidth]{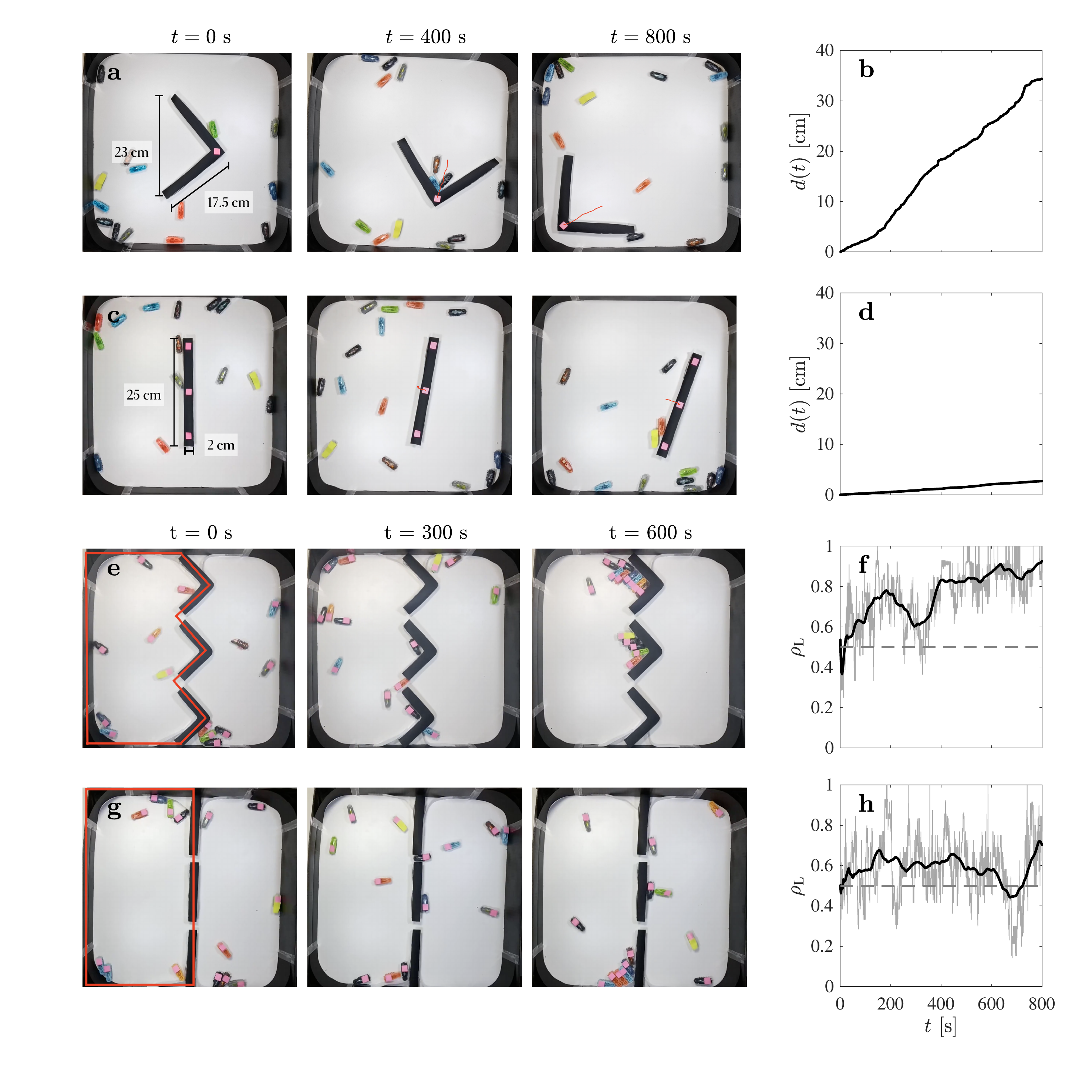}
    \caption{
    {\bf Interactions with objects and obstacles.}
    {\bf a} The active particles present in the arena push a wedge (black V-shaped object with the pink tag). The orange line represents its trajectory in the past 30 seconds (see Video~4).
    {\bf b} Its corresponding displacement grows approximately linearly in time at a rate of $\approx 0.038\,{\rm cm\,s^{-1}}$.
    {\bf c} The active particles do not push as efficiently a more symmetric linear object (see Video~5),
    {\bf d} as demonstrated by the fact that its displacement growth rate is about one order of magnitude smaller.
    {\bf e} In the presence of a series of fixed wedges dividing the arena, the active particles get gradually segregated to the left of the wedges (area within the red line, see Video~6),
    {\bf f} as demonstrated by the growth of the probability $\rho_{\rm L}$ of the active particles to be on the left side of the arena as a function of time. (The gray line is the frame-by-frame value. The black line is a smoothed version created by calculating the moving average with a span of 10\% of the data points).
    {\bf g} If the obstacles are symmetric, the active particles remain uniformly distributed across the entire arena (see Video~7),
    {\bf h} as demonstrated by the fact that $\rho_{\rm L}$ remains close to $0.5$ for the entire duration of the experiment.
    }
    \label{fig:2}
\end{figure*}

\section{Interaction Between Objects and Obstacles} \label{obstacles}

We now consider the effect of the interaction of active particles with movable objects and fixed obstacles present in their environment.
In the case of movable objects, the presence of active particles generates an active bath that displaces such objects.
If these objects are asymmetric, they can then feature directed motion or rotation.
In the case of fixed obstacles, these can be used to alter the way in which the active particles explore their space.

\subsection{Movable Rods and Wedges} 

Active particles interact with a movable object by pushing it \cite{angelani2010geometrically,mallory2014curvature,kaiser2014transport,knezevic2020effective,solon2022einstein}. If the object is asymmetric, it will be propelled because the active particles interact with it asymmetrically.
For example, when we placed a wedge in the arena with 14 Hexbugs (Fig.~\ref{fig:2}a and Video~4),
we observed that the Hexbugs tended to get trapped on its concave side and push the wedge. Meanwhile they tended to interact for much shorter times with its convex side, because they slid along the wedge without getting trapped.
We tracked the tip of the wedge using a pink tag and measured its total displacement as a function of time. This is shown by the black line in Fig.~\ref{fig:2}b, which grows approximately linearly in time at a rate of $\approx 0.038\,{\rm cm\,s^{-1}}$

As a control experiment, we considered also a linear rod (Fig.~\ref{fig:2}c and Video~5). 
In this case, the Hexbugs interacted with both sides of the rod in an equal manner, resulting overall in much less movement.
In fact, when the Hexbugs enter into contact with the rod, they align with it and slide along its length without pushing it too much.
This is quantified by the fact that its displacement growth rate, shown in Fig.~\ref{fig:2}d, is about one order of magnitude smaller than that in the case of the wedge.

In the literature, there are several observations of passive particles interacting with an active bath. 
Several experiments showed that symmetric objects such as spheres or rods immersed in an active bath feature superdiffusion at short time scales and enhanced diffusion at long time scales \cite{wu2000particle}.
For elongated shapes, enhanced angular diffusion has also been observed \cite{yang2016dynamics,peng2016diffusion}.
There are also several experiments with asymmetric particles propelled in an active bath.
These include, for example, arrow-shaped tracers (referred to as ``shuttles'') \cite{angelani2010geometrically}, wedges \cite{kaiser2014transport}, concave shapes \cite{mallory2014curvature,knezevic2020effective}, and asymmetric dumbbells \cite{belan2021active}. 
In particular, Ref.~\cite{kaiser2014transport} demonstrated how the activity of motile bacteria in a solution pushed a wedge similarly to what we have shown in Fig.~\ref{fig:2}a and Video~4.

\subsection{Fixed Rods and Wedges}

We can also consider the case where there are fixed obstacles in the environment. In fact, because active particles are not in thermodynamic equilibrium with their environment, it is possible to use the features of the environment to perform complex tasks on the active particles. Some examples are separating, trapping, or sorting them based on their motion properties \cite{bechinger2016active}.
For example, as shown in Fig.~\ref{fig:2}e, we performed an experiment with a series of wedges fixed along the middle of the arena, with a gap between them of approximately $2\,{\rm cm}$, wide enough for the Hexbugs to go through. 
We then placed 14 Hexbugs randomly distributed in the arena and let them free to move.
As time passes (see Video~6), these wedge-shaped barriers effectively trap the active particles on the left side of the arena (area encircled by the red solid line in Fig.~\ref{fig:2}e). 
This is expected because the concave shape of the barrier rectifies the motion of the Hexbugs and either makes them turn around or traps them in its corners.
Fig.~\ref{fig:2}f shows the fraction $\rho_{\rm L}$ of active particles in the left portion of the box as a function of time. With our system parameters, the distribution quickly approaches a plateau of $\rho_{\rm L} \approx 0.9$.

Also in this case, we performed a control experiment using rods instead of wedges as obstacles, as shown in Fig.~\ref{fig:2}g.
In this case, the Hexbugs remain uniformly distributed throughout the arena (see Video~7).
This is quantified by measuring the fraction of active particles on the left side of the arena (area encircled by the red solid line in Fig.~\ref{fig:2}g). As shown in Fig.~\ref{fig:2}h, the distribution hovers at $\rho_{\rm L} \approx 0.5$ for the whole duration for the experiment.

Similar behaviors have been observed with microscopic particles. 
For example, Ref.~\cite{galajda2007wall} showed that {\it E. coli} bacteria can be concentrated by a wall of funnels to the subspace to where the funnel openings lead (like those in Fig.~\ref{fig:2}e). 
They also repeated the experiment with motile bacteria using a flat wall with openings (as in Fig.~\ref{fig:2}g) finding that, in this case, the concentration remains the same, on average, in the two subspaces. 

\begin{figure*}
    \includegraphics[width=.8\linewidth]{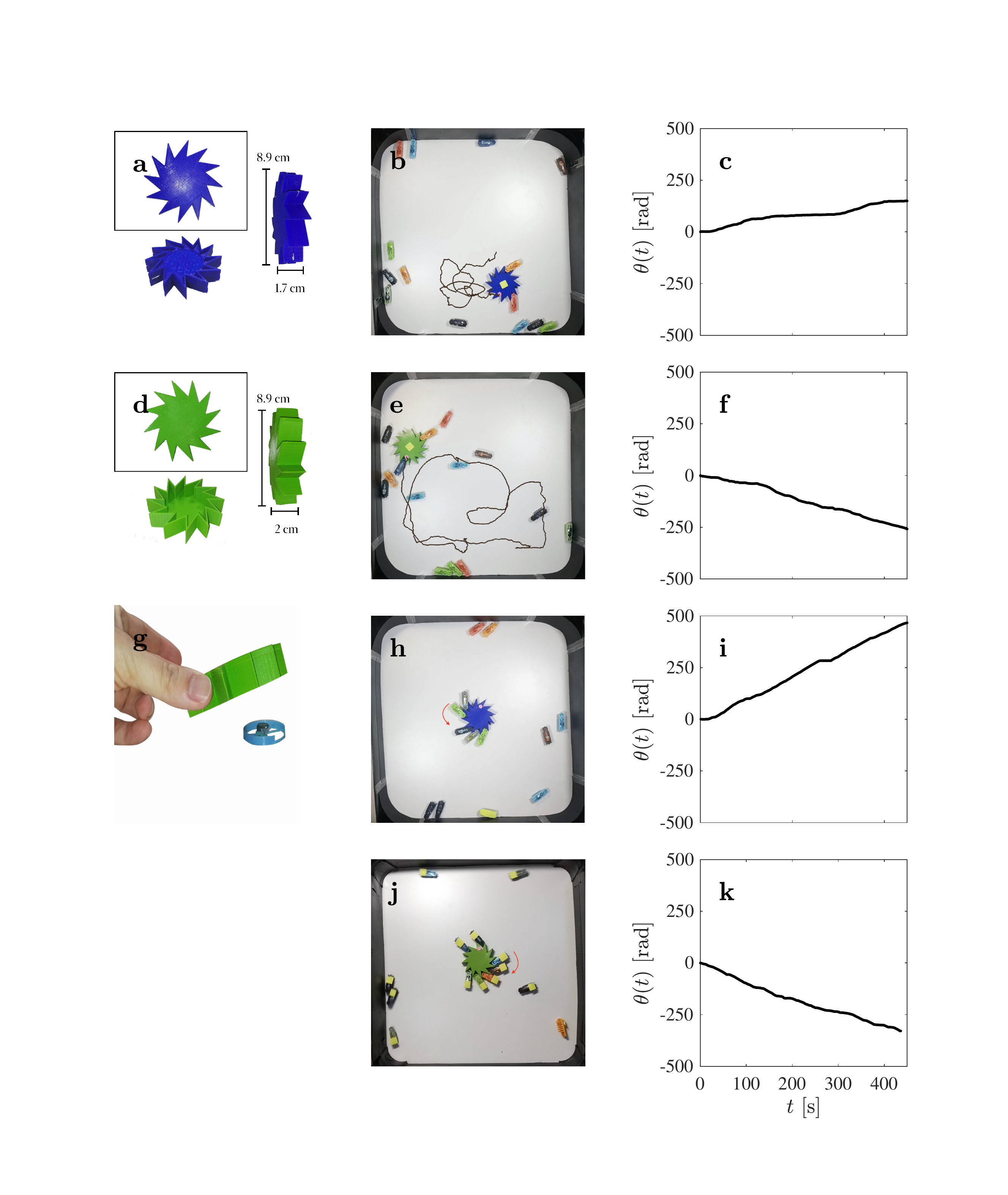}
    \caption{
    {\bf Interaction with gears.}
    {\bf a} An asymmetric gear is an obstacle with a chiral shape.
    {\bf b} When it is placed in the arena with the active particles, it is pushed and made to rotate (see Video~8). The trajectory corresponds to 30 seconds.
    {\bf c} The orientation of the gear $\theta(t)$ grows at a rate of $\approx +0.7\,{\rm rad\,s^{-1}}$.
    {\bf d}-{\bf f} The gear with the opposite chirality rotates in the opposite direction with a rate $\approx -1.1\,{\rm rad\,s^{-1}}$ (see Video~9).
    {\bf g} The position of the gear can then be fixed to center of the arena by pinning it on a fixed pivot around which it is still free to rotate.
    {\bf h}-{\bf i} The gear rotation is enhanced to $\approx +2.3\,{\rm rad\,s^{-1}}$ because now all the forces exerted by the active particles contribute to its rotation and not to its translation (see Video~10).
    {\bf j}-{\bf k} Similarly, the rotation of the opposite-chirality gear is enhanced but in the opposite direction to $\approx -1.5\,{\rm rad\,s^{-1}}$ (see Video~11).
    }
    \label{fig:3}
\end{figure*}

\subsection{Movable Gears}

To demonstrate that it is also possible to transfer torque to objects, we set up new experiments replacing the movable wedge (and rod) with asymmetric gears. In this context, we refer as gears to small disk-shaped objects with several saw teeth. We show an example of such a gear in Fig.~\ref{fig:3}a. 
This gear has multiple concave corners created by the teeth. When active particles, like Hexbugs, interact with these gears, they behave similarly as in the case of the V-shaped objects. However, in this case, the propulsion of the Hexbugs also causes the gear to rotate.

In the experiment, we placed the gear in the middle of an arena with 14 Hexbugs (Fig. \ref{fig:3}b and Video~8). 
We tracked the motion of the gear by placing two bright-colored tags on top of it: one on the center and the other on one tooth.
The active particles tend to get trapped in the saw teeth and push the gear around. However, because the saw teeth are located all around the gear, the propelling force also creates a torque. Therefore the gear not only moves linearly, but also rotates in a  direction that depends on the orientation of its teeth.
To quantify this behavior, we analyzed the trajectory and measured the cumulative angle of rotation (Fig.~\ref{fig:3}c). As expected, the data shows that the gear rotates in the left-chiral direction. The rate of the rotation is $\approx +0.7\, {\rm rad}\cdot{\rm s}^{-1}$.

To compare, we ran a second experiment with another gear, which has the same size, but whose teeth have the opposite orientation (Fig. \ref{fig:3}e and Video ~8). By performing the same analysis as in the previous experiment, we determined that the behavior of the active particles was similar. However, the gear rotates in the right-chiral direction (Fig. \ref{fig:3}f). The rate of the rotation is $\approx -1.1\, {\rm rad}\cdot{\rm s}^{-1}$.

These behaviors were first predicted with numerical simulations in Ref.~\cite{angelani2009selfstarting}. They were then experimentally realized using microscopic gears set in active baths of motile {\em B. subtilis} \cite{sokolov2009swimming} and {\em E. coli} \cite{dileonardo2010bacterial} bacteria. The dimensions of the gears used in the experiments were in the range of 10s to 100s of micrometers and the achieved angular speeds of the order of a few revolutions per minute (rpm).
Interestingly, these works demonstrated that it is in principle possible to extract energy from the disordered motion of an active bath by obtaining directed rotational motion.
There have also been alternative proposals to obtain directed rotational motion by using catalytic self-propelling Janus particles \cite{maggi2016selfassembly} and by generating the torque using thermocapillary forces \cite{maggi2015micromotors} or self-electrophoresis \cite{brooks2019shapedirected}.

\subsection{Fixed Gears}

We finally explore the behavior of the system when the gears are fixed to a surface. To achieve this, we glued a metallic nut to the center of the arena. As the bottom of the gear is hollow, we could place them on top of the nut (Fig.~\ref{fig:3}g). This setup restrains the position of the gear while still enabling it to rotate.
As in the previous experiment, we let 14 Hexbugs move freely within the arena (Fig.~\ref{fig:3}h and Video ~10). We observe that the Hexbugs get trapped in the teeth of the gear, inducing its right-chiral rotation. By analyzing the motion of the gear, we measured the rotation rate $\approx +2.3\, {\rm rad\,s^{-1}}$ (Fig.~\ref{fig:3}h). 
When compared to the data shown in Fig.~\ref{fig:3}c, we see that the rotation rate is enhanced when the gear is fixed to the surface. This is expected because, as the linear displacement is restrained, the propelling force is applied to the gear exclusively as a torque, which increases the rotation speed.

To compare, we performed a second experiment with the other gear, which has the opposite chirality (Fig.~\ref{fig:3}j and Video ~11). As expected, the ensuing behavior was similar, but displaying left-chiral rotation (Fig.~\ref{fig:3}h). The gear rotates at a rate of approximately $-1.5\, {\rm rad\,s^{-1}}$. Also in this case, we see that the rotation is enhanced with respect to the case of the freely moving gear in Fig.~\ref{fig:3}f.

A microscopic version of this experiment was performed in Ref.~\cite{vizsnyiczai2016light}, where motile bacteria were used to power an array of microgears. 
The angular speed of each microgear depended on several factors, but it could reach up to $20\,{\rm rpm}$ with an edge speed very close to the speed of a freely swimming bacterium. 
This result showed that it is possible to exploit motile bacteria to rotate fixed microgears, like microscopic draught animal moving equally microscopic millstones. 

\begin{figure*}
    \includegraphics[width=0.8\linewidth]{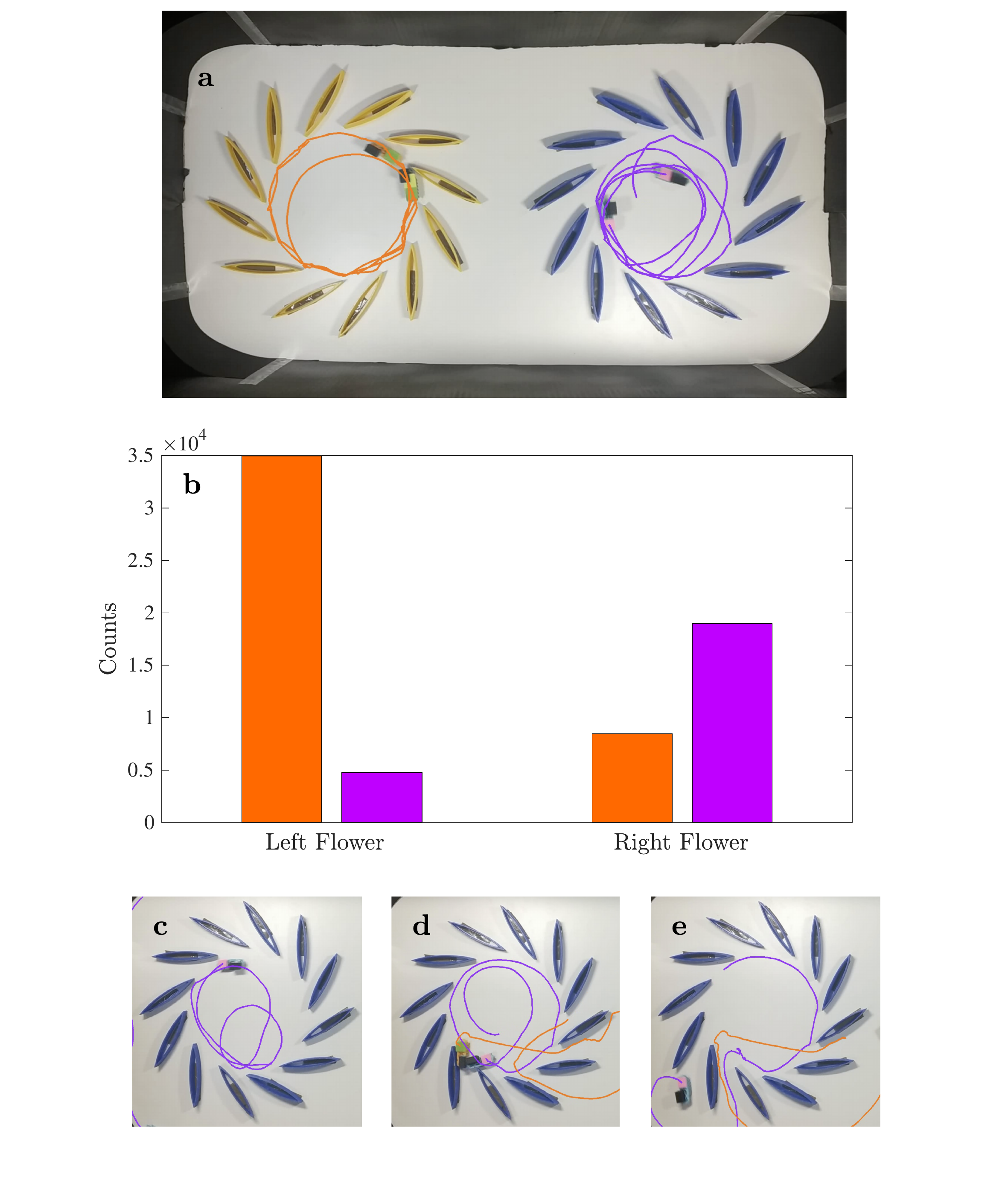}
    \caption{
    {\bf Sorting of chiral active particles.}
    {\bf a} Left-chiral (orange trajectories) and right-chiral (violet trajectories) active particles trapped in a left-chiral (yellow) and right-chiral (blue) flower, respectively. Each trajectory corresponds to 5 seconds (see Video~12).
    {\bf b} Number of counts for the two active particles in each of the chiral flowers.
    {\bf c}-{\bf e} Example of the interaction between a left-chiral active particle (orange trajectory) and a right-chiral active particle (violet trajectory) and a right-chiral flower:
    {\bf c} the right-chiral active particle is trapped in the right-chiral flower; 
    {\bf d} when the left-chiral particle arrives, the motion of the right-chiral active particle is disrupted;
    {\bf e} since the motion of the left-chiral active particle is not matched to the geometry of the right-chiral flower, it quickly moves out of the chiral flower.
    }
    \label{fig:4}
\end{figure*}

\section{Sorting of Chiral Active Particles}

In this section, we show that, by modifying the environment appropriately, we can sort chiral active particles according to their chirality.
To achieve sorting, we assemble a structure called a ``chiral flower''. This structure is made up of a set of ellipses placed equidistantly along a circumference, resembling ``petals'', tilted with respect to the radius of the circumference \cite{mijalkov2013sorting}.
We build an arena of $88\,{\rm cm}\times 44\,{\rm cm}$ and place two chiral flowers with an inner radius of $10\,{\rm cm}$ (Fig.~\ref{fig:4}a). The petals of one of the flowers have a tilt of $+60^{\circ}$, while those of the other flower have a tilt of $-60^{\circ}$. 
We then release 4 Hexbugs in the arena: 2 performing right-chiral (Fig.~\ref{fig:1}d) and 2 left-chiral (Fig.~\ref{fig:1}g) motion.

Throughout the experiment (Video~12), we observe that the left-chiral Hexbugs spend more time within the left-chiral flower (yellow), while the right-chiral Hexbugs spend more time within the right-chiral flower (blue).
We quantify this behavior by counting how many times each Hexbug is found within a specific flower. Fig.~\ref{fig:4}b shows that the left-chiral Hexbugs were found in the left-chiral flower seven times more frequently than the right-chiral ones and that the right-chiral Hexbugs were found two times more frequently in the right-chiral flower than the left-chiral ones.

Figs.~\ref{fig:4}c-e show a series of snapshots of the motion of a right-chiral Hexbug (violet trajectory) in a right-chiral flower: the bug remains trapped within the flower until a left-chiral Hexbug (orange trajectory) enters the right-chiral flower and disrupts its motion, pushing it out of the right-chiral flower.
The left-chiral Hexbug itself quickly moves out of the right-chiral flower because its motion is not matched to the geometry of the right-chiral flower.

Using the chiral flowers to sort chiral active particles was originally proposed in Ref.~\cite{mijalkov2013sorting}, where it was shown with numerical simulations that chiral flowers can sort active particles of different chiralities down to the size of large biomolecules.

\begin{figure*}
    \includegraphics[width=\linewidth]{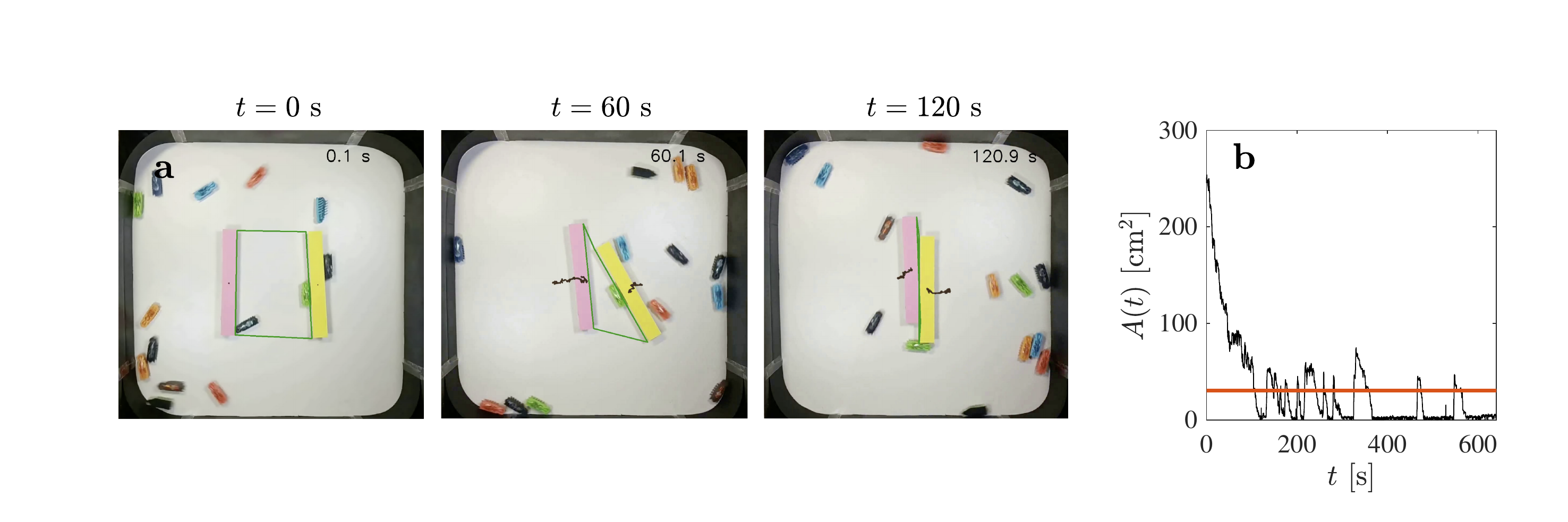}
    \caption{ 
    {\bf Casimir-like activity-induced forces between objects in an active bath.}
    {\bf a} Two straight rods (pink and yellow) are placed within an active bath, i.e., an arena where several active particles are present. As the active particles interact with the rods, the rods tend to be pushed together (see Video~13) --- an effect reminiscent of depletion and Casimir forces \cite{ray2014casimir}.
    {\bf b} This attraction can be quantified by measuring the decrease of the area $A(t)$ between the two rods (outlined by the green polygon in {\bf a} as a function of time. The orange solid line represent the area corresponding to the two rods being parallel at a distance equal to the width of a Hexbug.
    }
    \label{fig:5}
\end{figure*}

\section{Casimir-like activity-induced forces}

We finally demonstrate how objects in an active bath can experience some activity-induced attraction.
In fact, when the persistence length of the active particles is comparable to the characteristic size of the confining geometry, their intrinsic active nature can give rise to attractive force between the objects \cite{kjeldbjerg2021theory}.
This is reminiscent of the Casimir attraction experienced by parallel metallic plates set at a small distance as a consequence of the confinement of the fluctuations of the vacuum electromagnetic field \cite{casimir1948influence}. 
Another example of fluctuation-induced forces are the critical Casimir forces experienced by colloidal particles in a binary critical mixture close to a second-orded phase transition due to density concentration fluctuations \cite{fisher1978wall,callegari2021optical}.

To obtain Casimir-like activity-induced forces, we place two parallel, straight rods at a distance of about $15\,{\rm cm}$ in the arena where 14 Hexbugs are present.
In Fig.~\ref{fig:5}a and Video ~13, we see how, as time passes, the rods are pushed closer to each other.

We quantify this behavior by measuring the area between the rods, defined by a polygon connecting their corners (green polygons in Fig.~\ref{fig:5}a).
To calculate $A(t)$ we use the shoelace formula \cite{shoelace}: \begin{equation}
    A=\frac{1}{2} \left| \sum_{i=1}^{n} \left( x_i y_{i+1}-x_{i+1}y_{i}\right) \right|,
\end{equation}
where $x_n$ and $y_n$ are the Cartesian coordinates of the polygon, listed in clockwise order.
As shown in Fig.~\ref{fig:5}b, this area decreases from $A(t = 0\,{\rm s}) \approx 270\,{\rm cm^2}$ reaching $A(t)=0\,{\rm cm^2}$ at $t\approx 120\,{\rm s}$. 
Subsequently, occasionally $A(t)$ increases to a value $\approx 30\,{\rm cm^2}$ (given by the product of the Hexbug width $w_{\rm Hexbug} = 1.7\,\rm{cm}$ and the length of the rod $h_{\rm rod} = 18\,\rm{cm}$, indicated by the orange solid line in Fig.~\ref{fig:5}b), when a Hexbug opens a path between the rods, before going back to $\approx 0\,{\rm cm^2}$.

Similar behaviors have been observed with microscopic particles. 
Ref.~\cite{angelani2011effective} demonstrated the attractive action of a bacterial bath on a solution of suspended spherical particles with experiments and numerical simulations. 
Ref.~\cite{ray2014casimir} simulated the attraction between parallel plates caused by active particles. Using a minimal model for the active particles and their interaction with the plates, they showed that the attraction between the plates increases with increasing running length of the active particles. 
Ref.~\cite{kjeldbjerg2021theory} provides a theory for the interaction between parallel plates mediated by active particles. 
It is worth noting that both Ref.~\cite{ray2014casimir} and Ref.~\cite{kjeldbjerg2021theory} assume a parallel plane geometry and infinitesimally small particles, while in our experiment the parallelism of the rods is not enforced and the size of the active particles is finite. 
Ref.~\cite{harder2014role} investigated the role of the shape of the suspended objects in determining the forces mediated by active particles, assuming finite-size active colloids; they found that large, elongated objects, like our rods, typically interact attractively, as we observed in our experiment.

\section{Conclusions}

With the experiments we have presented in this article, we have shown that Hexbugs emulate the behavior of active particles. 
We have also shown how to characterize their active Brownian motion, estimating their propulsion speed, their angular diffusion, and their persistence length. 
Furthermore, they can also be modified to perform chiral active Brownian motion. 
We have shown that they can propel and rotate asymmetric objects, allowing to extract work from an active bath. 
With an appropriate design of the arena, we have shown that it is possible to concentrate the Hexbugs on one side using a separating panel in the form of a sieve, or sorting Hexbugs depending on their chirality using chiral flowers. 
We have also shown that the Hexbugs can induce attraction between two rods mimicking the Casimir effect mediated by active matter.

These small robots can be used to demonstrate nonequilibrium phenomena in a macroscopic scale, suitable for the classroom.
In fact, performing physical and tangible experiments, as opposed to using only simulations \cite{volpe2014simulation}, makes the learning process more intuitive.

Beyond the examples presented in this article, it is possible to use the Hexbugs to emulate many other experiments that have been performed with active matter, such as those described in Ref.~\cite{bechinger2016active}.
For example, it is possible to further modify the environment to create other interesting behaviors, e.g., by adding objects or obstacles of different sizes and shapes.
It is also possible to modify the Hexbugs to make them respond to some external stimuli, like light \cite{mijalkov2016engineering, leyman2018tuning} or sound intensity \cite{ahmed2015selectively,ahmed2016artificial,eliakim2018fully,dillinger2021ultrasound,ahmed2021bioinspired}.

\providecommand{\noopsort}[1]{}\providecommand{\singleletter}[1]{#1}%

\end{document}